\documentclass{article}
\usepackage{spconf,amsmath,graphicx}
\usepackage{booktabs}
\newcommand{\ra}[1]{\renewcommand{\arraystretch}{#1}}

\title{A Generative-First Neural Audio Autoencoder}
\name{Jonah Casebeer, Ge Zhu, Zhepei Wang, Nicholas J. Bryan}
\address{Adobe Research}

\begin{document}
\ninept
\maketitle

\begin{abstract}
Neural autoencoders underpin generative models. Practical, large-scale use of neural autoencoders for generative modeling necessitates fast encoding, low latent rates, and a single model across representations. Existing approaches are reconstruction-first: they incur high latent rates, slow encoding, and separate architectures for discrete vs. continuous latents and for different audio channel formats, hindering workflows from preprocessing to inference conditioning. We introduce a generative-first architecture for audio autoencoding that increases temporal downsampling from $2048\times$ to $3360\times$ and supports continuous and discrete representations and common audio channel formats in one model. By balancing compression, quality, and speed, it delivers $10\times$ faster encoding, $1.6\times$ lower rates, and eliminates channel-format-specific variants while maintaining competitive reconstruction quality. This enables applications previously constrained by processing costs: a 60-second mono signal compresses to 788 tokens, making generative modeling more tractable.
\end{abstract}


\section{Introduction}
Latent generative models have revolutionized audio synthesis, enabling applications from music generation~\cite{copet2023simple,evans2025stable,ning2025diffrhythm} to source separation~\cite{erdogan2023tokensplit}, upmixing~\cite{bralios2025learning}, and understanding tasks~\cite{goel2025audio}. 
These models fundamentally depend on neural audio autoencoders to compress raw waveforms into tractable latent representations. 
However, existing autoencoders are primarily designed as quantized variational autoencoders for reconstruction tasks, then mildly adapted for generative modeling through post-hoc modifications. This reconstruction-first design philosophy creates fundamental mismatches with generative requirements, leading to inefficient tokenization rates (we use ``tokenize'' to describe converting audio to any latent representation), fragmented architectures across audio channel formats, and computational bottlenecks that limit practical deployment at scale.
The discrete-continuous latent representation divide further compounds these issues: discrete methods lack continuous latents for diffusion, and continuous methods lack discrete tokens for language models.
These challenges are particularly acute for high-fidelity music processing at 44.1 kHz, where complex content amplifies the tension between compression, quality, and complexity.

\begin{figure}[t!]
    \centering
    \includegraphics[width=.925\linewidth, trim={0 0 0 0}, clip]{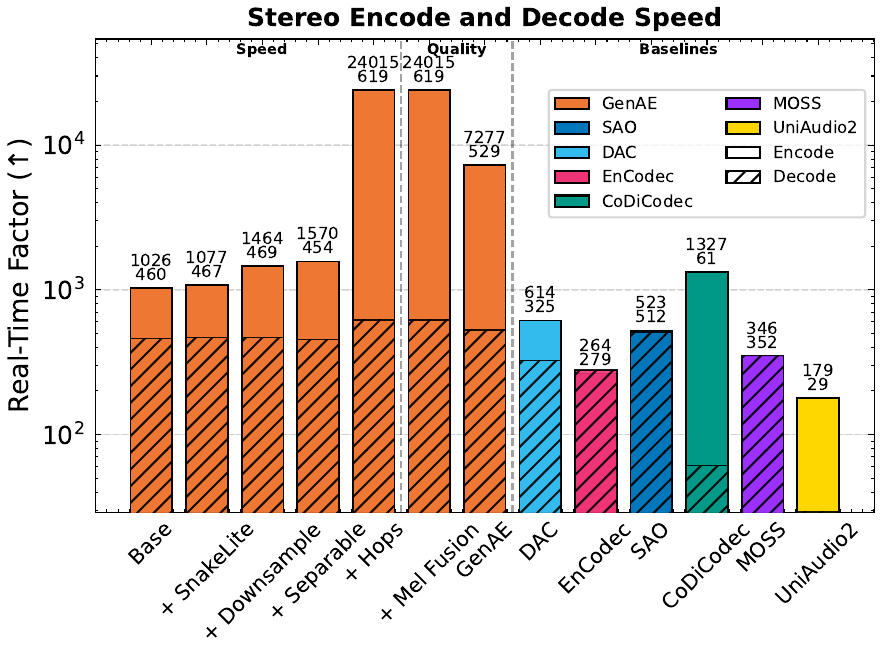}
    \vspace{-12pt}
    \caption{Encode log RTF (solid) and decode log RTF (striped) for GenAE ablations and baselines. GenAE modifications are split into ``speed'' and ``quality'' categories. Baselines are shown on the right. Faster encoding accelerates generative workflows.}
    \vspace{-4pt}
    \label{fig:speed_comparison}
\end{figure}

These limitations are evident across established audio codecs. SoundStream, EnCodec, and DAC~\cite{zeghidour2021soundstream, defossez2023high, kumar2023high} exemplify this paradigm, operating at 75-150 Hz with only quantized latents unsuitable for diffusion training. A 4-minute song requires over 18,000 tokens, creating memory bottlenecks, while slow encoding can constitute 30\% of training time, limiting data augmentation and throughput. Recent attempts achieve success with significant trade-offs: Stable Audio Open~\cite{evans2025stable} reduces rates to 21.5 Hz but increases encoding costs and only provides continuous latents unsuitable for language model training, SpectroStream~\cite{li2025spectrostream} maintains low rates but requires 64 codebooks, and HILCodec~\cite{ahn2024hilcodec} improves speed but maintains high token rates. The Music2Latent~\cite{pasini2024music2latent} line of work, culminating in CoDiCodec~\cite{pasini2025codicodec}, achieved impressive 11 Hz compression rates along with continuous and discrete latents, at the cost of performance in signal-level metrics. None of these explicitly account for different audio channel formats. Speech-focused approaches~\cite{casanova2025nanocodec, parkerscaling, defossez2024moshi} achieve 12.5 Hz but target low-bandwidth applications unsuitable for high-fidelity music with its broader frequency content and complex stereo imaging. These developments point toward the need for unified architectures designed specifically for generative modeling. Concurrent works~\cite{yang2026uniaudio, gong2026moss} push music to 12.5 Hz on 24 kHz audio.

To address these limitations, we propose the Generative-First Autoencoder (GenAE), a generative-first architecture that rethinks previous autoencoder designs for generation. GenAE provides a single architecture and training scheme that supports continuous and discrete latents, and all common audio channel formats (mono, stereo, mid/side). Although we ablate and evaluate on 44.1 kHz music, the design is not music-specific. Our contributions are threefold: (1) encoder architectural modifications including efficient activations, early downsampling, and strategic attention placement that enable aggressive tokenization with substantial computational speedups, (2) training improvements with audio channel format data augmentation and loss functions that enhance generalization and robustness, and (3) an optional post-training step that discretizes a trained continuous model to support both continuous and discrete latents, without retraining the backbone. Together, these choices yield a unified model that balances compression rate, reconstruction quality, and processing speed for generative workflows.

\section{Method}
\begin{figure}[t!]
    \centering
    \includegraphics[width=0.72\linewidth, trim={20 35 20 35},clip]{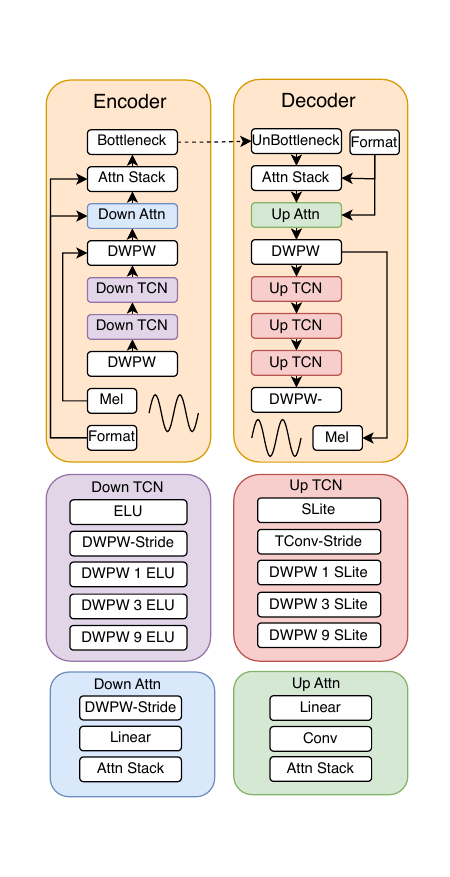}
    \vspace{-4pt}
    \caption{GenAE Model Architecture: DWPW is a standard depth-wise/point-wise layer with a dilation (1/3/9). Attn Stack is a standard multi-head windowed attention block. SLite is the SnakeLite activation. TCN is a standard dilated residual convolution block. Mel represents inputting or outputting a mel-spectrogram. Format represents an audio channel format token.}
    \vspace{-4pt}
    \label{fig:genae_arch}
\end{figure}

Our method redesigns the autoencoder architecture for generation to balance three competing objectives: compression rate, reconstruction quality, and processing speed. Our base model is an encoder-bottleneck-decoder type model, as proposed in SoundStream~\cite{zeghidour2021soundstream}. We specifically use the most modern architecture proposed in DAC~\cite{kumar2023high} with 5 blocks and hop sizes set for a target rate of 13.125 Hz. In this section, we describe each change to DAC to arrive at our proposed GenAE to ensure measurable and cumulative performance gains. We first sort our contributions into architectural, training, and post-training categories, and within each group by intent. A diagram depicting the cumulative model architecture is in Fig.~\ref{fig:genae_arch}.

\subsection{Architecture optimization}
The first four are for efficiency, ordered by measured impact, and the last three are for quality.

\noindent \textbf{Efficient activations:} Snake activations, $x + \frac{\sin^2(\beta x)}{\beta}$, excel in audio tasks~\cite{kumar2023high} but incur significant memory costs and are the memory bottleneck in our tasks. We use ELU in the encoder and introduce SnakeLite for the decoder. SnakeLite is the periodically wrapped Taylor approximation of $\sin^2(\cdot)$. We wrap the argument to Snake to $(-\pi/2,\pi/2]$ using the round function,
$a(x,\beta) = \beta x - \pi \, \operatorname{round}\!\left(\frac{\beta x}{\pi}\right)$, which we provide to the Taylor polynomial 
$P_8(z) = z^2 - \frac{z^4}{3} + \frac{2 z^6}{45} - \frac{z^8}{315} \approx \sin^2 z$,
and compose:
$$\text{SnakeLite}(x,\beta) = x + \frac{P_8\big(a(x,\beta)\big)}{\beta}$$

\noindent \textbf{Early downsampling:} We move downsampling from after to before the residual blocks in the encoder~\cite{howard2017mobilenets,tan2019efficientnet}. This reduces complexity from $O(L \cdot C^2)$ to $O(L/r \cdot C^2)$ where $r$ is the downsampling ratio in each block, $L$ is sequence length, and $C$ is the channel size. 

\noindent \textbf{Separable convolutions:} We replace dense convolutions with separable convolutions~\cite{howard2017mobilenets}. This reduces complexity from $O(C_{in} \cdot C_{out} \cdot K)$ to $O(C_{in} \cdot K + C_{in} \cdot C_{out})$. Separable transposed convolutions trigger a CUDA fallback to a slower GEMM call, so we omit them from the decoder.

\noindent \textbf{Aggressive temporal downsampling:} With more efficient layers, the next strategy is to reduce the number of layers. We combine downsampling layers in the encoder and reduce from 5 to 3. In the decoder, we perform a less aggressive decrease from 5 to 4. To compensate, we increase channel dimensions, hypothesizing that temporal compression can be offset by increased representational capacity. 

\noindent \textbf{Mel-spectrogram fusion:} Aggressive downsampling discards high-frequency information. We add an auxiliary pathway with explicit spectral content. We concatenate mel-spectrograms with intermediate encoder features, matching the mel hop size to the convolutional encoder downsampling rate. This allows the network to capture high-frequency content and low-frequency phase, without affecting speed. We add a symmetric mel output head to the decoder. 

\noindent \textbf{Windowed self-attention:} 
We strategically swap convolutional blocks for self-attention at four key representation bottlenecks: pre-final downsample, post-final downsample, post-bottleneck, and post-first upsample. These locations are the most compressed stages of the network. This approach provides maximum capacity where it can have the most impact.

\noindent \textbf{Unified multi-format conditioning:} To enable processing a variety of audio channel formats (mono, stereo, and mid/side), we introduce a conditioning mechanism via an audio channel format token. We create dedicated tokens for mid/side/left/right channels and embed each as a 64-dimensional learnable vector. The format tokens are injected into the attention layers using adaptive layer normalization.

\subsection{Training optimization}

\noindent \textbf{Unified multi-format augmentation:}
To enable multi-format inference, we use a corresponding multi-format training procedure. For each stereo sample in our dataset, we randomly select, with equal probability, to use a single channel, convert to mono, or convert to mid/side. We select the appropriate format embedding to condition the model and apply level augmentation across all formats.

\noindent \textbf{Auxiliary mel loss:}
We add an L1 mel-spectrogram reconstruction loss between the input mel features and the decoder's mel-head predictions. This auxiliary loss serves two purposes: it accelerates training convergence and encourages a division of labor within the network. It trains the convolutional layers to perform vocoding while guiding the attention layers to focus on higher-level processing.

\noindent \textbf{Co-prime multi-resolution losses:}
To mitigate harmonic bias, all STFT window sizes used in our reconstruction and discriminator losses are chosen to be coprime~\cite{schwar2023multi}. This ensures that the model cannot leverage shared windowing artifacts across all losses.

\subsection{Post-training optimization}
\noindent \textbf{Unified latent representations:} To support both continuous and discrete latents, we use a two-stage procedure within a single model. First, we train a continuous latent representation (GenAE-KL). We then apply latent restructuring~\cite{bralios2025learning, bralios2025re} to learn an inner RVQ bottleneck, discretizing the representation without retraining the backbone (GenAE-VQ). While post-hoc adaptation in the reverse direction (VQ$\to$KL) has been explored~\cite{zhu2025review}, we adopt the complementary KL$\to$VQ route for practical reasons: we and others have found that end-to-end VQ training at very low rates can be brittle and resource-intensive~\cite{parkerscaling, defossez2024moshi}, and using pre-quantizer latents from a discrete model can complicate diffusion training~\cite{zhu2025review}. Adding RVQ post-hoc leverages the stability of the continuous model and allows the same architecture to encode into and decode from either KL or VQ latents, providing flexibility for downstream applications and facilitating controlled comparisons between diffusion/flow models and language models.

\section{Experimental Setup}
We aim to evaluate GenAE's reconstruction quality, speed, and generative utility from three angles. First, a speed ablation across architectural updates to quantify the contribution of each design decision. Second, a benchmark against state-of-the-art baselines to compare the quality/compression trade-off across various bottlenecks and metrics. Third, a downstream-focused evaluation where we quantify the context enabled by a model and its flexibility in terms of audio channel format. We denote the continuous version as GenAE-KL and the discrete version as GenAE-VQ to distinguish their respective bottleneck types.

\subsection{Model details}
We train 13.125 Hz and 36.75 Hz GenAE variants with identical architectures but different compression rates. The 13.125 Hz GenAE uses DAC-style encoder blocks (32/64 channels) with 16$\times$/15$\times$ downsampling, mel-spectrogram fusion (192-bin, window 1792, hop 240), 3-layer transformers (512d, 2048 FFN, 8 heads), and 64d latents. The decoder uses 6-layer transformers (768d, 3072 FFN, 12 heads) and upsampling [15, 8, 2]. The 36.75 Hz GenAE modifies ratios to [15$\times$, 10$\times$] encoding, mel hop 150, 2/4 transformer layers, and [15, 5, 2] upsampling. Both use dropout 0.05, weight normalization, QK-normalization, RoPE, windowed attention (size 16), and 64d AdaLN conditioning.

\subsection{Training details}
We train on 25K hours of licensed instrumental stereo music at 44.1 kHz using 8 A100 GPUs for one week with AdamW ($1 \times 10^{-4}$, $\beta_1=0.8$, $\beta_2=0.9$), 1024-batch warmup, exponential decay (0.999999), and gradient clipping (norm 10). Our most comparable baseline, SAO, trains for 19 days on 32 A100 GPUs. We use loss weights: mel reconstruction 10, mel fusion 5, discriminator 1, feature matching 5. Batch size: 24 segments of 1.219 s. The continuous representation targets KL divergence 15. VQ uses Re-Bottleneck~\cite{bralios2025re} with 8-layer transformers (512d, 2048 FFN, 8 heads), 16 codebooks (1024 entries, 16d). Our Re-Bottleneck training for RVQ takes 4 days on 4 A100 GPUs at a batch of 64 with 4 s segments.

\subsection{Evaluation metrics} 
We measure processing speed via real-time factor (RTF) and reconstruction quality using scale-invariant signal-to-distortion-ratio (SI-SDR), multi-resolution short-time Fourier transform (STFT) loss, mel-spectrogram L1 distance, PESQ-WB, and Audiobox Aesthetics, all in bfloat16 precision. We choose bfloat16 over float32 or float64 as nearly all generative modeling is performed in reduced precision, and is our target application. Note that float32 yields significantly different values but similar trends. For baselines at non-44.1 kHz rates, we resample to native rates for processing and back to 44.1 kHz for evaluation. We use the Song Describer~\cite{manco2023song} dataset and filter out all vocals using a voice activity detection model. We only train on instrumental content and aim to only evaluate on instrumental content.

\subsection{Baseline comparisons}
We evaluate GenAE through speed, audio channel format, and latent format ablation studies. We compare with a suite of state-of-the-art codecs. Baselines include DAC, EnCodec (24/48 kHz), Stable Audio Open, CoDiCodec (discrete/continuous), MOSS, and UniAudio2.

\section{Results}
\begin{figure*}[t!]
    \centering
    \includegraphics[width=.925\linewidth]{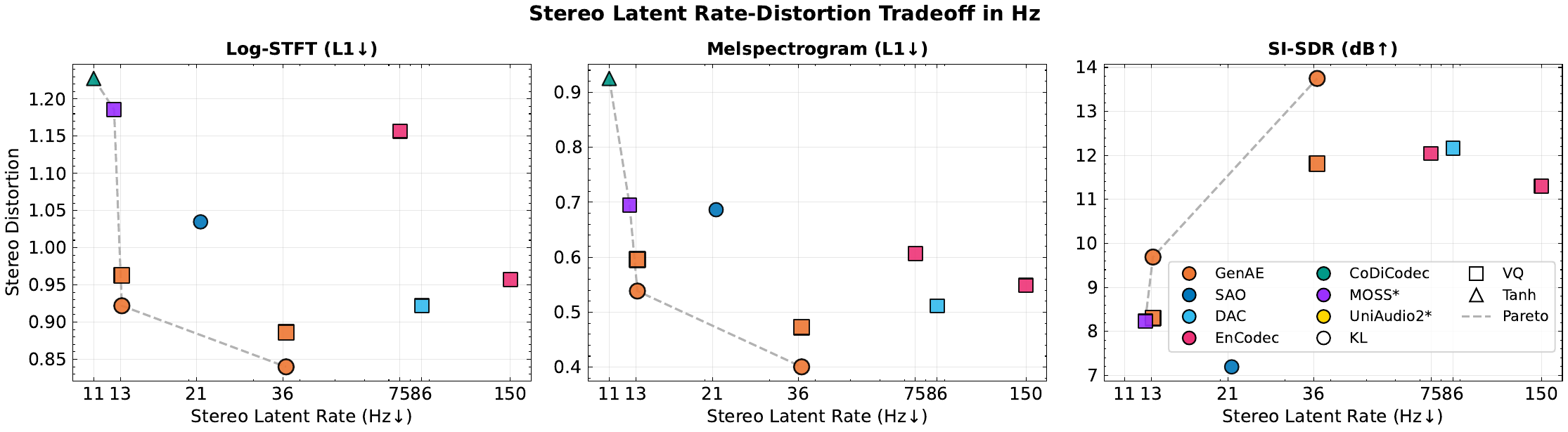}
    \vspace{-7pt}
    \caption{Stereo rate-distortion vs. latent rate (Hz). GenAE at 13 Hz matches baselines at far lower rates and at 36 Hz surpasses all baselines. Lower rates reduce tokens and memory for long-context generation. GenAE is Pareto-optimal on the compression/reconstruction frontier in both continuous (-KL) and discrete (-VQ) latent modes. PESQ-WB and Audiobox Aesthetics follow a similar trend. Models in the legend but not shown exceed plot ranges. Models with a * only run at 24 KHz. The Pareto frontier is shown in grey.}
    \vspace{-7pt}
    \label{fig:rd_hz}
\end{figure*}

\subsection{Speed benchmark}
Our generative-first design philosophy prioritizes encoding speed, recognizing that generative models require encoding massive datasets during training. This makes encoding speed key to training throughput and to on-the-fly data augmentations (e.g. stem remixing, pitch/time shifts) where precomputing all possible augmentation-combinations would be impractical. We evaluate design decisions using real-time factor (processed/elapsed time). Fig.~\ref{fig:speed_comparison} shows results for encoding and decoding 2 60-second stereo samples at 44.1 kHz. Starting from our DAC-like baseline with Snake activations (Base), we achieve progressive improvements through: efficient activations (+5\% encode), pre-residual downsampling (+36\% encode), separable convolutions (+7\% encode), and aggressive temporal downsampling (15$\times$ encoding speedup). Additional components include mel-spectrogram fusion and windowed attention with format conditioning. The resulting GenAE model uses 3$\times$ less memory than SAO, encodes over 10$\times$ faster than SAO, and decodes 1.6$\times$ faster than DAC, enabling faster processing for easier generative modeling.

\subsection{Compression and quality benchmark}
Effective generative modeling requires autoencoders that achieve aggressive compression and good reconstruction quality while supporting continuous and discrete latents, enabling diffusion or language modeling. Our evaluation demonstrates how GenAE achieves this joint objective. Fig.~\ref{fig:rd_hz} shows GenAE achieving superior rate-distortion performance essential for generative modeling. At 13.125 Hz, GenAE surpasses SAO in all metrics while running at 60\% of the rate, and matches DAC performance at just 15\% of the rate. This directly enables faster training and reduced memory requirements for generative models. At 36.75 Hz, GenAE-KL outperforms all baselines across all metrics while maintaining substantial compression. GenAE-VQ at 13.125 Hz again outperforms SAO across all metrics, demonstrating unified architecture performance across both continuous and discrete representations. Note that CoDiCodec sounds better than the metrics show. This unification eliminates the need for separate models when comparing diffusion and language model approaches, addressing a benchmarking challenge~\cite{tal2025auto}. We evaluate PESQ-WB at 16 kHz following~\cite{lanzendorfer2025evaluating}. GenAE-KL maintains its rate-distortion frontier position, with 36.75 Hz achieving the best score (4.04) and 13.125 Hz providing strong high-compression performance (3.00). Baselines score: DAC (3.49), EnCodec-48 (3.77), EnCodec-24 (3.67), SAO (2.76), MOSS (2.82), CoDiCodec (1.73), UniAudio2 (1.31). We conduct reference-free evaluation using the Audiobox Aesthetics model~\cite{tjandra2025meta}, which reveals tighter clustering. GenAE-KL scores 7.87 (36.75 Hz) and 7.88 (13.125 Hz), while baselines range from 7.77-7.88: MOSS (7.88), EnCodec-48 (7.83), SAO (7.82), EnCodec-24 (7.80), and CoDiCodec/DAC/UniAudio2 (7.77-7.78). Both metrics operate at 16 kHz, which may favor lower-bandwidth models like EnCodec-24, MOSS, and UniAudio2.

\subsection{Multi-format unification benchmark}
Practical generative modeling requires autoencoders that work seamlessly across different audio channel configurations. Our unified architecture eliminates the need for multiple specialized models, reducing deployment complexity with consistent performance across major audio channel formats. Tab.~\ref{tab:format_mono_side} evaluates mel-spectrogram reconstruction performance across L/R and M/S stereo audio channel formats. GenAE performs nearly identically in both -KL and -VQ forms across audio channel formats, demonstrating consistent quality regardless of input audio channel format or internal latent configuration. This is crucial for generative models to process diverse content. In contrast, CoDiCodec, SAO and EnCodec-48 degrade significantly due to L/R-specific training, while mono models are varied (DAC/UniAudio2/MOSS maintain, EnCodec-24 degrades). This highlights the importance of multi-format design since GenAE-based models can manipulate different audio formats without needing to re-encode/decode.

\subsection{Generative context benchmark}
To illustrate the practical benefits of aggressive compression for generative modeling, we estimate maximum context lengths on a 40 GB GPU at bfloat16 with a batch of 8 and 80\% VRAM budget for activations. For language models, we assume a MusicGen-like~\cite{copet2023simple} model with 24 attention layers, 2048 dimensions, and a KV cache. For diffusion models, we assume 24 attention layers and 1024 dimensions. Tab.~\ref{tab:format_mono_side} shows the resulting maximum context capacity. At 13.125 Hz, GenAE-VQ enables 832 s of context, 6.5$\times$ longer than the next-best model, DAC. For continuous diffusion modeling, GenAE-KL at 13.125 Hz achieves 173 s of context, compared to SAO's 106. CoDiCodec provides superior compression enabling $12\%-20\%$ more context. We note that GenAE is the only model where additional compression does not come with the tradeoff of increased encoding time. Recent benchmarking~\cite{tal2025auto} suggests that higher compression rates enable improved downstream modeling performance.

\setlength{\tabcolsep}{1.15pt}
\begin{table}[h!]
    \centering
    \ra{.85}
    \caption{Long-context memory use and multi-format performance benchmark. GenAE has consistent performance across latent layouts and audio channel formats, enabling long-form generative modeling.}
    \begin{tabular*}{\linewidth}{@{\extracolsep{\fill}}l c c c c@{}}\toprule
        \textbf{Model} & \textbf{Hz}$\downarrow$ & \textbf{Context (s)}$\uparrow$ & \textbf{L/R mel}$\downarrow$ & \textbf{M/S mel}$\downarrow$ \\ \midrule
        EnCodec-24 & 75 & 146 & 0.6062 & 0.6383 \\
        EnCodec-48 & 150 & 73 & 0.5485 & 0.6601 \\
        DAC & 86 & 127 & 0.5114 & 0.5144 \\
        CoDiCodec-FSQ & 11 & 993 & 0.9596 & 1.0558 \\
        MOSS & 12.5 & 873 & 0.6945 & 0.6958 \\
        UniAudio2 & 12.5 & 873 & 2.0371 & 2.0561 \\
        GenAE-VQ (ours) & 13.125 & 832 & 0.5956 & 0.5943 \\
        GenAE-VQ (ours) & 36.75 & 297 & 0.4726 & 0.4766 \\ \midrule
        SAO & 21.5 & 106 & 0.6863 & 0.7506 \\
        CoDiCodec & 11 & 206 & 0.9251 & 1.0217 \\
        GenAE-KL (ours) & 13.125 & 173 & 0.5384 & 0.5369 \\
        GenAE-KL (ours) & 36.75 & 62 & 0.4005 & 0.4054 \\
        \bottomrule
    \end{tabular*}
    \vspace{-4pt}
    \label{tab:format_mono_side}
\end{table}

\section{Conclusion}
In this work, we design neural audio autoencoders for generative models. Our GenAE is a single backbone that autoencodes at rates as low as 13.125 Hz ($3360\times$ downsampling), supports both continuous (KL) and discrete (VQ) latents, and operates across mono, stereo, and mid/side audio channel formats without model variants. A principled combination of efficient activations, early downsampling, separable convolutions, mel-spectrogram fusion, and windowed attention yields order-of-magnitude faster encoding and substantially lower memory while maintaining competitive fidelity, placing GenAE on the rate-distortion frontier with far fewer tokens. Our generative-first design repositions tokenization from a computational bottleneck to an efficient, scalable component: a 60-second track compresses to 788 tokens, enabling tractable long-context transformers. A post-training step lets us use both continuous and discrete latents in one model.

\bibliographystyle{IEEEbib}
\bibliography{refs}

@article{zeghidour2021soundstream,
  title={Soundstream: An end-to-end neural audio codec},
  author={Zeghidour, Neil and Luebs, Alejandro and Omran, Ahmed and Skoglund, Jan and Tagliasacchi, Marco},
  journal={IEEE/ACM Transactions on Audio, Speech, and Language Processing},
  year={2021}
}

@article{defossez2023high,
  title={High Fidelity Neural Audio Compression},
  author={D{\'e}fossez, Alexandre and Copet, Jade and Synnaeve, Gabriel and Adi, Yossi},
  journal={Trans. Mach. Learn. Res.},
  year={2023}
}

@article{kumar2023high,
  title={High-fidelity audio compression with improved {RVQGAN}},
  author={Kumar, Rithesh and Seetharaman, Prem and Luebs, Alejandro and Kumar, Ishaan and Kumar, Kundan},
  journal={Advances in Neural Information Processing Systems (NeurIPs)},
  year={2023}
}

@article{ahn2024hilcodec,
  title={{HILCodec}: High-Fidelity and Lightweight Neural Audio Codec},
  author={Ahn, Sunghwan and Woo, Beom Jun and Han, Min Hyun and Moon, Chanyeong and Kim, Nam Soo},
  journal={IEEE Journal of Selected Topics in Signal Processing},
  year={2024}
}

@article{casanova2025nanocodec,
  title={{NanoCodec}: Towards High-Quality Ultra Fast Speech LLM Inference},
  author={Casanova, Edresson and Neekhara, Paarth and Langman, Ryan and Hussain, Shehzeen and Ghosh, Subhankar and Yang, Xuesong and Juki{\'c}, Ante and Li, Jason and Ginsburg, Boris},
  journal={arXiv:2508.05835},
  year={2025}
}

@inproceedings{parkerscaling,
  title={Scaling Transformers for Low-Bitrate High-Quality Speech Coding},
  author={Parker, Julian D and Smirnov, Anton and Pons, Jordi and Carr, CJ and Zukowski, Zack and Evans, Zach and Liu, Xubo},
  year={2025},
  booktitle={International Conference on Learning Representations (ICLR)}
}

@article{defossez2024moshi,
  title={Moshi: a speech-text foundation model for real-time dialogue},
  author={D{\'e}fossez, Alexandre and Mazar{\'e}, Laurent and Orsini, Manu and Royer, Am{\'e}lie and P{\'e}rez, Patrick and J{\'e}gou, Herv{\'e} and Grave, Edouard and Zeghidour, Neil},
  journal={arXiv:2410.00037},
  year={2024}
}

@article{li2025spectrostream,
  title={SpectroStream: A Versatile Neural Codec for General Audio},
  author={Li, Yunpeng and Han, Kehang and McWilliams, Brian and Borsos, Zalan and Tagliasacchi, Marco},
  journal={arXiv:2508.05207},
  year={2025}
}

@article{pasini2024music2latent,
  title={Music2latent: Consistency autoencoders for latent audio compression},
  author={Pasini, Marco and Lattner, Stefan and Fazekas, George},
  journal={arXiv:2408.06500},
  year={2024}
}

@article{goel2025audio,
  title={{Audio Flamingo 3}: Advancing audio intelligence with fully open large audio language models},
  author={Goel, Arushi and Ghosh, Sreyan and Kim, Jaehyeon and Kumar, Sonal and Kong, Zhifeng and Lee, Sang-gil and Yang, Chao-Han Huck and Duraiswami, Ramani and Manocha, Dinesh and Valle, Rafael and others},
  journal={arXiv:2507.08128},
  year={2025}
}

@article{bralios2025re,
  title={{Re-Bottleneck}: Latent Re-Structuring for Neural Audio Autoencoders},
  author={Bralios, Dimitrios and Casebeer, Jonah and Smaragdis, Paris},
  journal={arXiv:2507.07867},
  year={2025}
}

@article{bralios2025learning,
  title={Learning to Upsample and Upmix Audio in the Latent Domain},
  author={Bralios, Dimitrios and Smaragdis, Paris and Casebeer, Jonah},
  journal={arXiv:2506.00681},
  year={2025}
}

@article{erdogan2023tokensplit,
  title={Tokensplit: Using discrete speech representations for direct, refined, and transcript-conditioned speech separation and recognition},
  author={Erdogan, Hakan and Wisdom, Scott and Chang, Xuankai and Borsos, Zal{\'a}n and Tagliasacchi, Marco and Zeghidour, Neil and Hershey, John R},
  journal={arXiv:2308.10415},
  year={2023}
}

@article{copet2023simple,
  title={Simple and controllable music generation},
  author={Copet, Jade and Kreuk, Felix and Gat, Itai and Remez, Tal and Kant, David and Synnaeve, Gabriel and Adi, Yossi and D{\'e}fossez, Alexandre},
  journal={Advances in Neural Information Processing Systems (NeurIPs)},
  year={2023}
}

@inproceedings{evans2025stable,
  title={Stable audio open},
  author={Evans, Zach and Parker, Julian D and Carr, CJ and Zukowski, Zack and Taylor, Josiah and Pons, Jordi},
  booktitle={IEEE International Conference on Acoustics, Speech and Signal Processing (ICASSP)},
  year={2025}
}

@article{ning2025diffrhythm,
  title={{DiffRhythm}: Blazingly fast and embarrassingly simple end-to-end full-length song generation with latent diffusion},
  author={Ning, Ziqian and Chen, Huakang and Jiang, Yuepeng and Hao, Chunbo and Ma, Guobin and Wang, Shuai and Yao, Jixun and Xie, Lei},
  journal={arXiv:2503.01183},
  year={2025}
}

@article{howard2017mobilenets,
  title={{MobileNets}: Efficient convolutional neural networks for mobile vision applications},
  author={Howard, Andrew G and Zhu, Menglong and Chen, Bo and Kalenichenko, Dmitry and Wang, Weijun and Weyand, Tobias and Andreetto, Marco and Adam, Hartwig},
  journal={arXiv:1704.04861},
  year={2017}
}

@inproceedings{tan2019efficientnet,
  title={{EfficientNet}: Rethinking model scaling for convolutional neural networks},
  author={Tan, Mingxing and Le, Quoc},
  booktitle={International Conference on Machine Learning (ICML)},
  year={2019}
}

@article{schwar2023multi,
  title={Multi-scale spectral loss revisited},
  author={Schw{\"a}r, Simon and M{\"u}ller, Meinard},
  journal={IEEE Signal Processing Letters},
  year={2023}
}

@article{tal2025auto,
  title={Auto-Regressive vs Flow-Matching: a Comparative Study of Modeling Paradigms for Text-to-Music Generation},
  author={Tal, Or and Kreuk, Felix and Adi, Yossi},
  journal={arXiv:2506.08570},
  year={2025}
}

@article{zhu2025review,
  title={A Review on Score-based Generative Models for Audio Applications},
  author={Zhu, Ge and Wen, Yutong and Duan, Zhiyao},
  journal={arXiv:2506.08457},
  year={2025}
}

@article{pasini2025codicodec,
  title={{CoDiCodec}: Unifying Continuous and Discrete Compressed Representations of Audio},
  author={Pasini, Marco and Lattner, Stefan and Fazekas, George},
  journal={arXiv preprint arXiv:2509.09836},
  year={2025}
}

@article{lanzendorfer2025evaluating,
  title={Evaluating Objective Speech Quality Metrics for Neural Audio Codecs},
  author={Lanzend{\"o}rfer, Luca A and Gr{\"o}tschla, Florian},
  journal={arXiv:2511.19734},
  year={2025}
}

@article{manco2023song,
  title={The Song Describer Dataset: a Corpus of Audio Captions for Music-and-Language Evaluation},
  author={Manco, Ilaria and Weck, Benno and Doh, Seungheon and Won, Minz and Zhang, Yixiao and Bogdanov, Dmitry and Wu, Yusong and Chen, Ke and Tovstogan, Philip and Benetos, Emmanouil and others},
  journal={Workshop on Machine Learning for Audio, Advances in Neural Information Processing Systems (NeurIPs)},
  year={2023},
}

@article{tjandra2025meta,
  title={Meta audiobox aesthetics: Unified automatic quality assessment for speech, music, and sound},
  author={Tjandra, Andros and Wu, Yi-Chiao and Guo, Baishan and Hoffman, John and Ellis, Brian and Vyas, Apoorv and Shi, Bowen and Chen, Sanyuan and Le, Matt and Zacharov, Nick and others},
  journal={arXiv:2502.05139},
  year={2025}
}

@article{yang2026uniaudio,
  title={UniAudio 2.0: A Unified Audio Language Model with Text-Aligned Factorized Audio Tokenization},
  author={Yang, Dongchao and Wang, Yuanyuan and Chong, Dading and Liu, Songxiang and Wu, Xixin and Meng, Helen},
  journal={arXiv:2602.04683},
  year={2026}
}

@article{gong2026moss,
  title={MOSS-Audio-Tokenizer: Scaling Audio Tokenizers for Future Audio Foundation Models},
  author={Gong, Yitian and Chen, Kuangwei and Fei, Zhaoye and Yang, Xiaogui and Chen, Ke and Wang, Yang and Huang, Kexin and Chen, Mingshu and Li, Ruixiao and Cheng, Qingyuan and others},
  journal={arXiv:2602.10934},
  year={2026}
}

\end{document}